\documentclass{appolb} 

\usepackage{epsfig}

\usepackage{amsmath}
\usepackage{amssymb}
\usepackage{euscript}

\newcommand{\Ucal}{{\cal U}}
\newcommand{\Peu}{\EuScript{P}}
\newcommand{\Keu}{\EuScript{K}}

\newcommand{\veps}{\varepsilon}

\newcommand{\tG}{{\bf G}'}
\newcommand{\Section}[1]{Section~\ref{#1}}
\begin{document}



\title{  Historical and mathematical aspects of  iterative solutions for Monte Carlo 
simulations\thanks{Presented at the Cracow Epiphany Conference on LHC Physics,
 Cracow, Poland, 4-6 January 2008, 
Acta Phys. Polon. {\bf B 39} (2008) 1761.}\;\thanks{Supported in part  by the EU grant MTKD-CT-2004-510126
in partnership with the CERN Physics Department,
by  RTN European Programme, MRTN-CT-2006-035505 (HEPTOOLS, Tools and Precision Calculations for Physics Discoveries at  Colliders)
 and by the Polish Ministry of
Scientific Research and Information Technology grant No 620/E-77/6.PRUE/DIE 188/
2005-2008.
}}

\author{Z. W\c{a}s
\address{Institute of Nuclear Physics, Polish Academy of Sciences,\\
  ul. Radzikowskiego 152, 31-342 Cracow, Poland}}

\maketitle
\begin{abstract}
Over the last 25 years Monte Carlo programs were being developped in Cracow in the group guided by 
Prof. Stanislaw Jadach. Many of those programs became standard in their 
application domains. In the following let us review some aspects of such 
projects which were probably at the foundation of their success.
We will concentrate on mathematical aspects of their design and history of their construction.
It is rather difficult to cover 25 years of the research in a single talk. That is why, I have organized 
my presentation around Monte Carlo {\tt PHOTOS} but stressing its relation to other 
activities and projects often realized together with Prof. Jadach. 
Many of omitted aspects will find their way
into other perentations collected in this volume.

I will concentrate on issues related to phasespace parametrization and spin amplitudes as used 
in our Monte Carlo programs such as  {\tt MUSTRAAL},  {\tt TAUOLA} or  
{\tt KKMC} and their similarities and differences with respect to solution 
used in  {\tt PHOTOS}.

\noindent
{\bf Preprint   IFJ-PAN-IV-2008-4}
\end{abstract}

\PACS{11.15.Pg, 13.40.Ks, 12.38.Cy}


\section{Introduction}
One of the essential steps in the construction of any algorithm for multi-particle final states is the appropriate analysis of the phase space parametrization.
In the {\tt PHOTOS} Monte Carlo \cite{Barberio:1994qi} for multi-photon production, 
an exact phase space parametrization is embodied in an iterative algorithm, the details of which are best described in \cite{Nanava:2006vv}. Control of the 
distributions and relative size of sub-samples for distinct numbers of 
final state particles requires a precise knowledge of the matrix elements 
including virtual corrections as well.
In the {\tt KKMC} Monte Carlo, the phase space generation is different, 
but the necessity to control matrix elements is also essential \cite{kkcpc:1999,Jadach:2000ir}.

Iterative procedures for parts of amplitudes, which are at the foundation of exponentiation \cite{Jadach:2000ir,Yennie:1961ad} and structure functions \cite{Altarelli:1977zs,Gribov:1972ri,Gorshkov:1966ht,Skrzypek:1992vk,RichterWas:1985gi} were exploited for the sake of use in {\tt KKMC} Monte Carlo.
In particular the  description of dominantly $s$-channel processes $e^+e^- \to \nu_e \bar \nu_e \gamma \gamma $ where,  $t$-channel $W$-exchange diagrams with gauge boson couplings, contribute to
matrix elements provide an interesting example \cite{Was:2004ig}.
These studies were motivated by  practical reasons, but also pointed at quite astonishing properties of tree-level spin amplitudes, namely that they can be separated into gauge invariant parts in a semi-automated way, easy to apply in the Kleiss-Stirling methods \cite{Kleiss:1985yh,Jadach:1998wp}.

One could ask the question whether similar techniques can be used in QCD, whether they are  of any practical use, and in fact to which degree they were already included in previous publications. These questions will be discussed elsewhere \cite{vanHameren,vanHameren:2008dy}.
We will not elaborate on these points requiring good understanding of factorization in QCD. Instead let us point to old, but important for me 
ref.~\cite{Berends:1982ie}, where properties of factorization for cross section, visualize themselves in a fully differential environment, even though only 
for QED and at first order of perturbartion expansion.
For the sake of caution, let us  mention the existence of limitations in such 
strategies, if applied to parton shower applications beyond NLO \cite{Kleiss:1990jv}.

Our presentation is organized as follows.
In \Section{SecNotation} we will discuss different aspects of phase space parametrization,
as used in {\tt PHOTOS} Monte Carlo and how it compares to 
other prgrams. 
Discussion of 
approximations necessary 
to construct crude distributions is started in  \Section{SecNotation}. 
Presentation 
 of the form of first order cross section, 
matrix elements and  approximations which were 
essential for construction of the first version  of the program 
is given  in  \Section{SecCommAnti}. 
With all  material collected, we will  point in  \Section{SecSingProd}
to mathematical properties of elements
used in the project, which actually made it possible, even though their documentation was 
never of high priority until now.
The summary in \Section{SecSummary} closes the paper.
%

\section{Phase space\label{SecNotation}}
\begin{figure}
\begin{center}
                       \epsfig{file=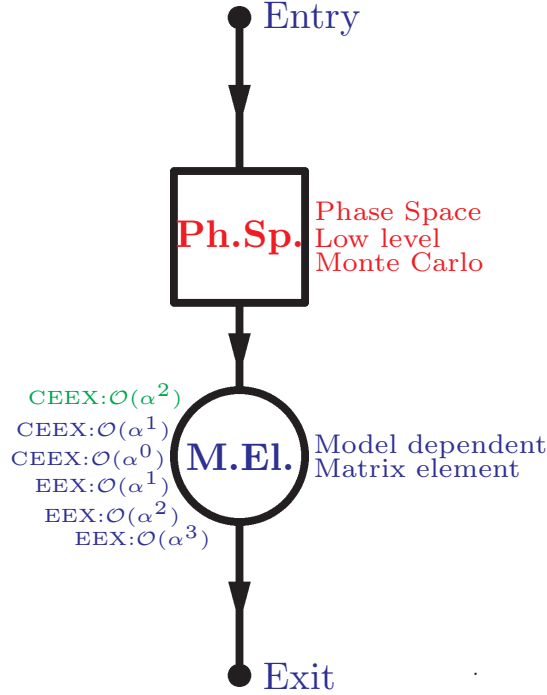,width=0.45\linewidth}
\caption{Phase space plot  for the KKMC and KORALZ Monte Carlo programs.\label{Fig1}}
\end{center}
\end{figure}

%
It is of no surprize that phase space must play a central role in preparation 
of the  algorithm of any Monte Carlo based on predictions originating from 
field theory. That is direct consequence of Quantum Mechanics,  basic formula
for cross section  consist of phase space element, matrix element squared and
the flux factor. Over many years we were stressing, in a multitude of 
talks and papers that the control of the eventual approximations is essential.
  Let me recall here one of such S. Jadach's  plots, see Fig. \ref{Fig1}.   
At that time it was an achievement\cite{koralz4:1994,kkcpc:1999}. It required enormous amount of work to
prepare such an organization of the phase space that would be
exact, cover complete multibody phase
space, and  capable to manage highly peaked distributions 
of complex structure due to collinear and soft singularities.

As these programs are discussed elsewhere in the proceedings,
 let us follow here the phase space
organization of another program originating from S. Jadach group, 
that is {\tt PHOTOS} Monte Carlo\footnote{ The most detailed decription 
of the program \cite{Barberio:1990ms,Barberio:1994qi},  
can be found  in recent ref.~\cite{Nanava:2006vv}.
}. It is also capable of covering multibody phase space 
distributions without any approximation, but contrary to {\tt KKMC/KORALZ} solutions
conformal symmetry of the eikonal approximation is not used. Thanks to that,
this solution is closer to iterative solution used in QCD parton showers, 
but is still relatively simple to explain and formalize.

 Let us start with the explicit expression for the parametrization of 
an  $n+1$ body phase space
in decay of  the object of four-momentum $P$\;  ($P^2=M^2$), as used in {\tt PHOTOS} Monte Carlo. 
As our aim is to define iterative relations, let us denote the four momenta 
of the first  $n$ decay products as $k_i$ ($i=1,n$) and the last $n+1$ decay product as $k_{n+1}$.
In our case  the $n+1$-th particle will  always be the real and massless  
photon\footnote{However the construction does not rely on a photon to be massless. 
In principle it can be 
applied to define other  phase space relations, for example the
emission of an extra massive pion or  emission of a pair of heavy particles.}. 
In the later steps of our construction the masslessnes of  photons and properties 
of QED matrix elements will be used. 

In the following, notation from refs. \cite{Was:1994kg,Jadach:1993hs} will be used. 
We will not rely on any particular results of these papers.
We only point to other,  similar options for
the exact $n$-body phase space parametrizations, which are also in use.

The Lorentz invariant phase space is defined as follows:
\begin{eqnarray}
dLips_{n+1}(P) &=&
{d^3k_1 \over 2k_1^0 (2\pi)^3}\; . . .\;{d^3k_n \over 2k_n^0 (2\pi)^3}
{d^3k_{n+1} \over 2k_{n+1}^0 (2\pi)^3}
(2\pi)^4 \delta^4\Bigl(P - k_{n+1}- \sum_{i=1}^n k_i\Bigr)\nonumber\\
&=&
d^4p\delta^4(P -p-k_{n+1}){d^3k_{n+1} \over 2k_{n+1}^0 (2\pi)^3}
{d^3k_1 \over 2k_1^0 (2\pi)^3} \;. . .\;{d^3k_n \over 2k_n^0 (2\pi)^3}
(2\pi)^4 \delta^4\Bigl(p -\sum_{i=1}^n k_i\Bigr)\nonumber\\
&=&
d^4p\delta^4(P -p-k_{n+1}){d^3k_{n+1} \over 2k_{n+1}^0 (2\pi)^3} dLips_n(p\to k_1 ... k_n),
\label{Lips_n+1}
\end{eqnarray}
 where extra  integration variables: four components of $p$ 
(compensated with $\delta^4\bigl(p -\sum_1^n k_i\bigr) $) is introduced.
 If further,    $M_{1...n}$ (compensated with 
$\delta\bigl(p^2 -M_{1...n}^2\bigr) $) is introduced,
the element of the phase space takes the form:
\begin{eqnarray}
dLips_{n+1}(P) &=&
{dM_{1...n}^2 \over (2\pi)} dLips_2(P \to p\ k_{n+1}) \times dLips_n(p \to k_1 ... k_n)\nonumber\\
&=&
dM_{1...n}^2  \biggl[d\cos\hat{\theta} d\hat{\phi} {1 \over 8(2\pi)^3}
{\lambda^{1\over 2}(M^2, {M_{1...n}^2 },{m_{n+1}^2 })\over M^2}\biggr]
\times dLips_n(p \to k_1 \dots k_n).\nonumber\\
\label{Lips_n+1.3}
\end{eqnarray}
The part of the phase space Jacobian corresponding to integration over the direction 
and energy of the last particle (or equivalently invariant mass $M_{1...n}$ of the remaining  system
of ${1...n}$ particles) 
is explicitly given.   
We will use later in the formulas   $m_i^2=k_i^2$, and analogously  $M_{i \dots n}$,
defining invariant masses of $k_i \dots k_n$ systems. 
The integration over the angles $\hat{\theta}$ and $\hat{\phi}$ is defined in the $P$ rest-frame. 
The integration over the invariant mass, $M_{1\dots n}$, is limited by phase space boundaries.
Anybody familiar with the phase space parametrization as used in {\tt FOWL} \cite{FOWL}, {\tt TAUOLA} \cite{Jadach:1993hs}, or many other programs will find the above explanation quite standard\footnote{%
%
The  parametrizations of such a type, 
use  properties of the Lorentz group in an explicit manner, in particular 
 measure,  representations and 
their products. That is why, they are  useful, for event building Monte Carlo 
programs   in 
phase space constructions based 
on boosts and rotations.
}. 

The question of choice of axes with respect to which angles are defined, and order 
in kinematical construction, is less trivial. The choice for the particular option stems from
 necessity to presample collinear singularities.  It is rather well known that the choice of 
the reference directions for the parametrization of the unit sphere is free, and can be used 
to advantage.
We will use related, but  somewhat different freedom of choice. Instead of  variables $\hat{\theta}\; \hat{\phi}$ defining orientation of  $k_{n+1}$ in $P$ rest-frame we will use  angles $\theta_1 \; \phi_1$
orienting $k_1$ (also  in $P$  rest-frame). The Jacobian for this reparametrization of unit sphere equals unity.

 Formula (\ref{Lips_n+1.3}) can 
be iterated and  provide a parametrization of the phase space with an arbitrary number of final state 
particles.  In such a case, the question of orientation of the frames used to define the angles
and the order of $M_{i\dots n}$ integrations (consequently, the choice of limits for $M_{i \dots n}$ integration), 
becomes particularly rich.
Our choice is defined in ref. \cite{Barberio:1994qi}.
We will not elaborate on this  point here.

If the invariant mass $M_{1\dots n}$ is replaced with the photon energy defined in the $P$ rest-frame, $k_\gamma$,
then the phase space formula can be written as:
\begin{eqnarray}
 dLips_{n+1}(P) &=&
\biggl[ k_\gamma dk_\gamma  d\cos\hat{\theta} d\hat{\phi} {1 \over 2(2\pi)^3}
\biggr]
\times dLips_n(p \to k_1 ... k_n),
\label{Lips_n+1.5}
\end{eqnarray}
If we would have $l$  photons accompanying $n$ other particles,
then the factor in square brackets is iterated.
The statistical factor ${1 \over l!}$ would complete the form of the phase space 
parametrization, similar to the exponent.
The last formula, supplemented with definition of frames with 
respect to which angles are defined  is used
to define the full kinematic configuration of the event. From angles and energies ($k_{\gamma_i})$ 
of photons and also angles, energies and masses of other decay products, 
four-momenta of all final state particles can be constructed.

If in formula (\ref{Lips_n+1.5}) instead of  $dLips_n(p \to k_1 ... k_n)$  one would use 
 $dLips_n(P \to k_1 ... k_n)$  the {\it \bf tangent space} would be obtained.  
Then $k_{n+1}$ photon does not affect  
 other particles' momenta at all, and thus has no boundaries on energy or direction. If this
formula would be iterated then
all such photons would be  independent from one another as well\footnote{ Expression (\ref{Lips_n+1.5}) would be  slightly more complicated if instead of photons
a massive particle was to be added.}.
Energy and momentum constraints on the photon(s) are introduced with  the relation between tangent 
and real $n+1$-body phase space. The formula defining one step in the iteration reads 
as follows\footnote{The $ \{ \bar k_1,\dots,\bar k_{n}\}$ can be identified with the event before 
the radiation of $k_\gamma$ is introduced.}:
\begin{eqnarray}
 && dLips_{n+1}(P\to k_1 ...  k_n,k_{n+1})=  dLips_{n}^{+1\; tangent} \times W^{n+1}_n, \nonumber\\[3mm]
&&dLips_{n}^{+1\; tangent} = dk_\gamma d\cos\theta d\phi \times dLips_n(P \to \bar k_1 ... \bar k_n),
\nonumber \\
&&\{k_1,\dots,k_{n+1}\} = {\bf T}\bigl(k_\gamma,\theta,\phi,\{\bar k_1,\dots,\bar k_n\}\bigr).
\label{Jacobians}
\end{eqnarray}
  The 
 $W^{n+1}_n$ depends on details of ${\bf T}$, and will be thus 
given later in formula~(\ref{Wnn}). To justify (\ref{Jacobians}), 
we have to convolute formula (\ref{Lips_n+1.3})
for $Lips_{n+1}(P \to  k_1 ... k_n,k_{n+1})$  with itself (for $Lips_{n}(p \to k_1 ... k_n)$):
\begin{eqnarray}
Lips_{n+1}(P \to k_1 ... k_n,k_{n+1}) &=& {dM_{1\dots n} ^2 \over 2\pi}  Lips_{2}(P \to k_{n+1} p) \times Lips_{n}(p \to  k_1 ... k_n) \nonumber \\ 
Lips_{n}(p \to k_1 ... k_n) &=& {dM_{2\dots n}^2 \over 2\pi}  Lips_{2}(p \to k_1 p') \times Lips_{n-1}(p' \to  k_2 ... k_n)
\label{AA}
\end{eqnarray}
and use it also for $Lips_{n}(P \to  \bar k_1 ... \bar k_n)$:
\begin{eqnarray}
Lips_{n}(P \to  \bar k_1 ... \bar k_n) &=& {dM_{2\dots n}^2 \over 2\pi}  Lips_{2}(P \to \bar k_1 \bar p') \times Lips_{n-1}(\bar p' \to  \bar k_2 ... \bar k_n).
\label{BB}
\end{eqnarray}

 Note that our  tangent space of variables $ dk_\gamma  d\cos{\theta} d{\phi}$ is unbounded from above and the limit is introduced
by $W_n^{n+1}$ which is set to zero for the configuations outside the phase sace. 
In principle, we should distinguish between variables like $M_{2\dots n} $ for invariant mass
of $k_2 \dots k_n$ and  $\bar M_{2\dots n} $ for invariant mass
of $\bar k_2 \dots \bar k_n$, but in our choice for $G_n$, $G_{n+1}$ below,  
$M_{2\dots n}= \bar M_{2\dots n}$ and  $M_{1\dots n} $ is defined anyway for the $n+1$-body phase space only.

 We direct the reader to refs.\cite{Barberio:1990ms,Barberio:1994qi} 
for an alternative presentation.  Let us remark that 
formula (\ref{Jacobians}) is quite general, many options, motivated by the
properties of the matrix elements, can be introduced. Generally the
transformation $T$ may differ from the choice to choice quite a lot. 
The most straightforward choice can be based on any $n$ and $n+1$ body phase space 
parametrizations using invariant masses and angles (e.g. exactly as in {\tt TAUOLA}
\cite{Jadach:1993hs} formulas 11 to 13).

If 
\begin{equation}
G_n \; : \; M_{2\dots n} ^2,\theta_{1},\phi_{1}, M_{3\dots n} ^2,\theta_{2},\phi_{2}, \dots, \theta_{n-1},\phi_{n-1} \;   \to \;\bar k_1 \dots \bar k_n 
\label{G-1}
\end{equation}
and 
\begin{equation} 
G_{n+1} \;:  \; k_\gamma,\theta,\phi,M_{2\dots n}^2,\theta_{1},\phi_{1}, M_{3\dots n} ^2,\theta_{2},\phi_{2},\dots, \theta_{n-1},\phi_{n-1} \;   \to \;k_1 \dots k_n,k_{n+1}
\label{G-2}
\end{equation}
then
\begin{equation}
{\bf T}=G_{n+1}( k_\gamma,\theta,\phi,G_n^{-1}(\bar k_1,\dots,\bar k_n)).
\end{equation}
The ratio of the Jacobians (factors $\lambda^{1/2}$  like in  formula (\ref{Lips_n+1.3}),
 etc.) form the factor $W^{n+1}_n$, which in our case is  rather simple, 
\begin{equation} 
W^{n+1}_n=  {k_\gamma}  {1 \over 2(2\pi)^3}
 \times \frac{\lambda^{1/2}(1,m_1^2/M_{1\dots n}^2,M^2_{2 \dots n}/M_{1\dots n}^2)}{\lambda^{1/2}(1,m_1^2/M^2,M^2_{2\dots n}/M^2)},
\label{Wnn}
\end{equation}
because of  choice for $G$ as explained in the Appendix of ref.\cite{Nanava:2006vv}. Note that 
${k_\gamma}=\frac{M^2-M_{1\dots n}^2}{2M}$.
There are additional benefits from such a choice. In all relations
$\bar k_2= Lk_2$, ...,  $\bar k_n= Lk_n$ and $\bar p'= Lp'$
common Lorentz transformation
 $L$ is used.     Transformation  $L$ 
is defined by $k_1,\bar k_1,\bar p',p'$ and $P$;
 internal relations between four vectors  $k_2 ... k_n$,
($\bar k_2 ... \bar k_n$) are not needed.

Formula (\ref{Jacobians}) can be realized algorithmically in the following way:
\begin{enumerate}
\item
 For any point in n-body phase space (earlier generated event), described for example with the explicit
configuration of four vectors $\bar k_1 ... \bar k_n$,
 coordinate variables can be calculated, using formula (\ref{G-1}).
\item Photon  variables can be generated according to Eq. (\ref{Jacobians}). The weight $W^{n+1}_n$ has to be 
also attributed.
\item Variables obtained in this way from the old configuration and the one of a  photon 
can be used to 
construct the new kinematical configuration for the $n+1$-body final state. 
The phase space weight, which is zero for configurations outside phase space 
boundaries, can be calculated at this point from (\ref{Jacobians},\ref{Wnn}) and finally combined  with
the matrix element.
\end{enumerate}

 Here we have chosen  two sub-groups of 
particles. The first one consisted of particle 
  1 alone, and the second, of particles 2 to n combined together.
 Obviously in the case of 2-body decays, 
there is not much choice when construction of the first photon is performed.

By iteration, we can generalize formula (\ref{Jacobians}) to the case of $l$  photons and
we write:
\begin{eqnarray}
 && dLips_{n+l}(P\to  k_1 ...  k_n, k_{n+1} ...  k_{n+l})= 
\frac{1}{l!} \prod_{i=1}^l  \biggl[ dk_{\gamma_i}  d\cos\theta_{\gamma_i} d\phi_{\gamma_i} 
 W^{n+i}_{n+i-1}\biggr]
\times dLips_n(P \to \bar k_1 ... \bar k_n),  \nonumber\\ &&
\{k_1,\dots,k_{n+l}\} = {\bf T}\bigl(k_{\gamma_l},\theta_{\gamma_l},\phi_{\gamma_l},{\bf T}\bigl( \dots,
{\bf T}\bigl(k_{\gamma_1},\theta_{\gamma_1},\phi_{\gamma_1},\{\bar k_1,\dots,\bar k_n\}\bigr) \dots\bigr).
\label{barred}
\end{eqnarray}
In this formula we can easily localize the   { \bf tangent space} for the multiple 
photon configuration.  In this space, each photon is independent from  
other particles' momenta.  Note that   
it is also possible  to fix upper boundary on $k_{\gamma_i} $ arbitrary high.
Photons are independent one from another as well. 
Correlations appear later,  thanks to iterated  transformation {\bf T}.
  The factors  $ W^{n+i}_{n+i-1}$ are calculated when 
constraints on each consecutive photon are introduced; the previously constructed ones 
are included in  the $n+i-1$ system\footnote{Configurations of $k_{\gamma_i} $ which can 
not be resolved are reduced to  the ones with that photon dropped out.}.

Of course, for the tangent space to be useful, the choice
of the definition of {\bf T} must be restricted at least by the condition $\{ k_1, \cdots k_n \} \to 
\{ \bar k_1, \cdots \bar k_n \}$ if all $k_{\gamma_i} \to 0$.\footnote{In fact further constraints have to be fulfilled to enable presampling for the collinear
singularities.
Note that  variables 
$k_{\gamma_m},\theta_{\gamma_m},\phi_{\gamma_m}$ are used at a time
of the $m-$th step of iteration  only, and are not needed elsewhere
in construction of the physical phase space; the same is true for invariants
and angles  $M_{2\dots n} ^2,\theta_{1},\phi_{1} ,\dots, \theta_{n-1},\phi_{n-1} \;   \to \;\bar k_1 \dots \bar k_n $ of (\ref{G-1},\ref{G-2}), which are also
redefined at each step of the iteration.
}

It is important to realize that  one has to  choose matrix elements on the tangent 
space to complete the construction used in {\tt PHOTOS}. The number and energies 
of photons will be generated on the tangent space first.
 Regularization of (at least) soft singularity must be defined.
Rejection, and event construction, is performed with the help of formula
(\ref{Jacobians})  for each consecutive photon. It  diminishes photon 
multiplicity with respect to the one defined for the tangent space.
 Of course, as rejection implements changes in phase space density, 
a matrix element (with  virtual corrections) of the physical space 
can be introduced  as well.

The treatment of the phase space presented here lies at the heart of the construction 
of {\tt PHOTOS} kinematics, and was used since its beginning. It exhausts the case when there is
only one charged particle in final state.  For multiple 
charged particle final states new complication appear, because  all
collinear configurations need simultaneous attention, and not only 
the one along $k_1$ direction.
A presampler with multichannel generation is needed.
In our case we follow the same method
 as explained  in ref. \cite{Jadach:1993hs}.

Let us now sum the above expression over $l$.
If we add arbitrary factors $f(k_{\gamma_i},\theta_{\gamma_i},\phi_{\gamma_i})$ and sum over $l$ we obtain:

\begin{eqnarray}
 &&  {\sum_{l=0}  \exp(-F)
\frac{1}{l!} \prod_{i=1}^l f(k_{\gamma_i},\theta_{\gamma_i},\phi_{\gamma_i})  } 
dLips_{n+l}(P\to  k_1 ...  k_n, k_{n+1} ...  k_{n+l})= \nonumber\\ &&
{ 
\sum_{l=0} \exp(-F)
\frac{1}{l!} \prod_{i=1}^l  }  
\biggl[ {
f(k_{\gamma_i},\theta_{\gamma_i},\phi_{\gamma_i})dk_{\gamma_i}  d\cos\theta_{\gamma_i} d\phi_{\gamma_i} }
 W^{n+i}_{n+i-1}\biggr]\times \nonumber\\ && dLips_n(P \to \bar k_1 ... \bar k_n),  \\ &&
\{k_1,\dots,k_{n+l}\} = {\bf T}\bigl(k_{\gamma_l},\theta_{\gamma_l},\phi_{\gamma_l},{\bf T}\bigl( \dots,
{\bf T}\bigl(k_{\gamma_1},\theta_{\gamma_1},\phi_{\gamma_1},\{\bar k_1,\dots,\bar k_n\}\bigr) \dots\bigr), \nonumber \\ &&
F  =\int_{k_{min}}^{k_{max}} dk_{\gamma} d\cos\theta_{\gamma} d\phi_{\gamma}f(k_{\gamma},\theta_{\gamma},\phi_{\gamma}). \nonumber
\label{barred0}
\end{eqnarray}

 Some  parts of rhs. taken  alone,  give 
crude distribution over tangent space (orthogonal set of variables $k_i,\theta_i, \phi_i$). 
 Factors $f$ must be integrable over this tangent space and  
regulators of singularities must be introduced.
We may simply request that 
\begin{eqnarray}
 &&  { \sigma_{tangent}} = 1= \nonumber\\ &&
{  
\sum_{l=0} \exp(-F)
\frac{1}{l!} \prod_{i=1}^l  }  
\biggl[ {
f(k_{\gamma_i},\theta_{\gamma_i},\phi_{\gamma_i})dk_{\gamma_i}  d\cos\theta_{\gamma_i} d\phi_{\gamma_i} } \biggr]
 \nonumber
\label{barred1}
\end{eqnarray}
and that sum rule originating from perturbative approach (Kinoshita-Lee-Nauenberg theorem) can be used
to control  virtual corrections; both for tangent and later also final 
distributions. 

At this point  we already have  Monte Carlo solution of {\tt PHOTOS} phase space.
In reality,  for that solution to work, real emission and virtual corrections 
need to be calculated and their factorization properties must be understood. 
That is why, choice of $f$ is free only in principle, in practice it
must be synchronized with those results for the sake of program efficiency.
In case of final state QED bremsstrahlung it is rather simple, eventual 
complications due to QED corrections to rates are of no major consequences 
\cite{Golonka:2006tw} for the program construction. Non leading corrections
appear only.

Note that this formula is very close to  other ones, used in other progams 
or calculations. For example formal solution 
\cite{RichterWas:1985gi,Skrzypek:1992vk}
of evolution equation reads
\begin{equation}
D(x,\beta_{ch})=\delta(1-x) + \beta_{ch}P(x) 
+ \frac{1}{2!}\beta_{ch}^2 \{P \times P \}(x) 
+ \frac{1}{3!}\beta_{ch}^3 \{P \times P \times P \}(x) + \dots
\end{equation}
where 
$P(x)=\delta(1-x)(\ln\varepsilon + 3/4) + \Theta(1-x - \varepsilon)
\frac{1}{x}(1+x^2)/(1-x)$ and 
$\{ P \times P\}(x) = \int_0^1 dx_1\int_0^1 dx_2 \delta(x-x_1 x_2) P(x_1)
P(x_2)$. 

One can easily observe, that in 
 the LL contributing regions, the  phase space Jacobian's as used in 
{\tt PHOTOS}
trivialize  \cite{Barberio:1994qi} and lead directly to
this solution. In 1994, this solution was truncated to second order. It was indeed 
profitable that solutions for similar problems were available in Cracow at that time.
Let us give one example \cite{Jadach:1987ii}. In this first,  on
multiphoton Monte Carlos, paper written 
in 1987 by  S. Jadach  formula (3.1)  is basically the same as
 tangent space of multi-photon {\tt PHOTOS} (and not much different from $D(x,\beta_{ch}$)\; discussed just above):
{
\begin{eqnarray}
\sigma(K)&  =exp \Bigl( \frac{2\alpha}{\pi}(\ln\frac{s}{m^2} -1) \ln\frac{k_s}{E} 
   +  \frac{\alpha}{\pi} \ln\frac{s}{m^2}\Bigr) \hskip 2 cm \nonumber \\
& \sum_{n=0} \frac{1}{n!} \prod_{m=1}^n \int_{k_s < k_m < K } \; \; \frac{d^3k_m}{k_m} 
\tilde S(k_1) \dots \tilde S(k_n) \tilde \beta_0
\end{eqnarray}
}
The difference appears in  projection from this tangent space to the physical 
one. Classical solution as proposed by Jadach,
use conformal symmetry,  projection  
from eikonal (tangent)  to physical space is performed in one step.
In  {\tt PHOTOS} eikonal symmetry is not used. Iterative 
 projection  is used instead, it is somewhat similar to the one introduced
 in {\tt TAUOLA} \cite{Jezabek:1991qp} for radiative corrections in leptonic tau decays.
Analogies to solutions used in QCD parton shower algorithms can be found.

Very important aspect of all these solutions is that the structure of singularities
is the same in tangent and final physical space. 
\section{Matrix elements\label{SecCommAnti}}
%
It is out of  question, that detailed 
analysis of {\tt MUSTRAAL}  Monte Carlo 
\cite{Berends:1982ie}, which was a consequence of accidental error in copying 
 source code from punch cards to tape, was essential for the design 
of {\tt PHOTOS} program. At that time (1983) I was forced to study 
{\tt MUSTRAAL} line 
after line.
Not only the two missing lines\footnote{Punch card reader glued them together
at the last time they were ever to be read?} of code were found, 
but I have studied the 
matrix element and crude distributions in all possible details.
This unintentionally collected experience combined with importance of QED 
radiative corrections in phenomenology of leptonic $Z$ couplings at the time
of preparation for first measurements of $\tau$ polarization at LEP
was few years later a starting point for {\tt PHOTOS}.

  Let us recall the properties of the $Z \to l^+ l^- \gamma$ matrix element
as studied by me at that early time and also the approximate matrix element,
 which was
and still is used in PHOTOS.

Let us write 
the explicit form  of the real-photon matrix element (separated from the phase space Jacobians),
for the $e^{+}e^{-} \to Z^{0}/\gamma^{*} \to \mu^{+}\mu^{-} (\gamma)$ process
 and as
used in the standard version of {\tt PHOTOS} (published in \cite{Barberio:1990ms,Barberio:1994qi}):
\begin{eqnarray}
X_{f}^{\mathrm{PHOTOS}}=&\frac{Q'^{2}\alpha(1-\Delta)}{4\pi^{2}s}s^{2} \hskip 3 mm \Bigg\{ \hskip 8 cm \nonumber \\
\frac{1}{k'_{+}+k'_{-}}\frac{1}{k'_{-}}&\bigg[(1+(1-x_{k})^{2})
\frac{{d}\sigma_{B}}{d\Omega}\Big(s,\frac{s(1-\cos\Theta_{+})}{2},
\frac{s(1+\cos\Theta_{+})
}{2}\Big)\bigg]\frac{(1+\beta\cos\Theta_{\gamma})}{2}\;\;\; \nonumber\\
+
\frac{1}{k'_{+}+k'_{-}}\frac{1}{k'_{+}}&\bigg[(1+(1-x_{k})^{2})
\frac{{d}\sigma_{B}}{d\Omega}\Big(s,\frac{s(1-\cos\Theta_{-})}{2},
\frac{s(1+\cos\Theta_{-})
}{2}\Big)\bigg]\frac{(1-\beta\cos\Theta_{\gamma})}{2}\Bigg\} \nonumber \\
\mathrm{where:} & \Theta_{+}=\angle(p_{+},q_{+}),\; \Theta_{-}=\angle(p_{-},q_{-}),
\;\hskip 4 cm \nonumber\\
 & \Theta_{\gamma}=\angle(\gamma,\mu^{-})\;  \textrm{is\, defined\,
in}\;(\mu^{+},\mu^{-})\textrm{-pair\, rest\, frame.} \hskip 1.2 cm
\label{X-fotos}
\end{eqnarray}
For its calculation (with respect to the Born cross-section)
it is enough to know the four momenta of the $Z$ and its decay products.
In the presented formulae we follow the notation from
refs.~\cite{Golonka:2006tw,Berends:1982ie}.
This expression  is to be compared with the exact one, taken from
ref.~\cite{Berends:1982ie}:
\begin{eqnarray}
X_{f}=\frac{Q'^{2}\alpha(1-\Delta)}{4\pi^{2}s}s^{2} &
\Bigg\{\frac{1}{(k'_{+}+k'_{-})}\frac{1}{k'_{-}}\bigg[\frac{{d}\sigma_{B}
}{{d}\Omega}(s,t,u')+\frac{{d}\sigma_{B}}{{d}\Omega}(s,t',u
)\bigg]\nonumber \\
 &
+\frac{1}{(k'_{+}+k'_{-})}\frac{1}{k'_{+}}\bigg[\frac{{d}\sigma_{B}}{{d}
\Omega}(s,t,u')+\frac{{d}\sigma_{B}}{{d}\Omega}(s,t',u)\bigg
]\Bigg\}.
\label{X-mustraal}
\end{eqnarray}

The resulting weight is rather simple, and reads:

\begin{eqnarray}
 WT_1=&  \frac{\frac{{d}\sigma_{B}
}{{d}\Omega}(s,t,u')+\frac{{d}\sigma_{B}}{{d}\Omega}(s,t',u
)}{\bigg[(1+(1-x_{k})^{2})
\frac{{d}\sigma_{B}}{d\Omega}\Big(s,\frac{s(1-\cos\Theta_{+})}{2},
\frac{s(1+\cos\Theta_{+})
}{2}\Big)\bigg]\frac{(1+\beta\cos\Theta_{\gamma})}{2}\; \big(1+ \frac{3}{4} \frac{\alpha}{\pi}\big)},    \hskip 5 cm    \nonumber \\
 WT_2=&   \frac{\frac{{d}\sigma_{B}}{{d}
\Omega}(s,t,u')+\frac{{d}\sigma_{B}}{{d}\Omega}(s,t',u)}{\bigg[(1+(1-x_{k})^{2})
\frac{{d}\sigma_{B}}{d\Omega}\Big(s,\frac{s(1-\cos\Theta_{-})}{2},
\frac{s(1+\cos\Theta_{-})
}{2}\Big)\bigg]\frac{(1-\beta\cos\Theta_{\gamma})}{2}\; \big(1+ \frac{3}{4} \frac{\alpha}{\pi}\big)}. \hskip 5 cm
\label{wgt1}
\end{eqnarray}

For its calculation the numerical
value of the electroweak couplings of $Z$ to fermions, as well as information on the state
from which the $Z$ was produced is nonetheless necessary. This seemingly trivial requirement puts
new stress on the event record: the details of the process of the $Z$ production need to be coded
in the event record, then correctly deciphered by {\tt PHOTOS} to calculate the process-dependent
weight. From our experience this requirement of {\tt  PHOTOS} may be difficult to accept by
other users of event records. The authors of event generators often choose their own conventions
in encoding the details of hard process such as  $q \bar q \to ng Z/\gamma^*; Z/\gamma^* \to \mu^+ \mu^-$
into the event record.

\begin{figure}
{\small
{ \resizebox*{0.49\textwidth}{!}{\includegraphics{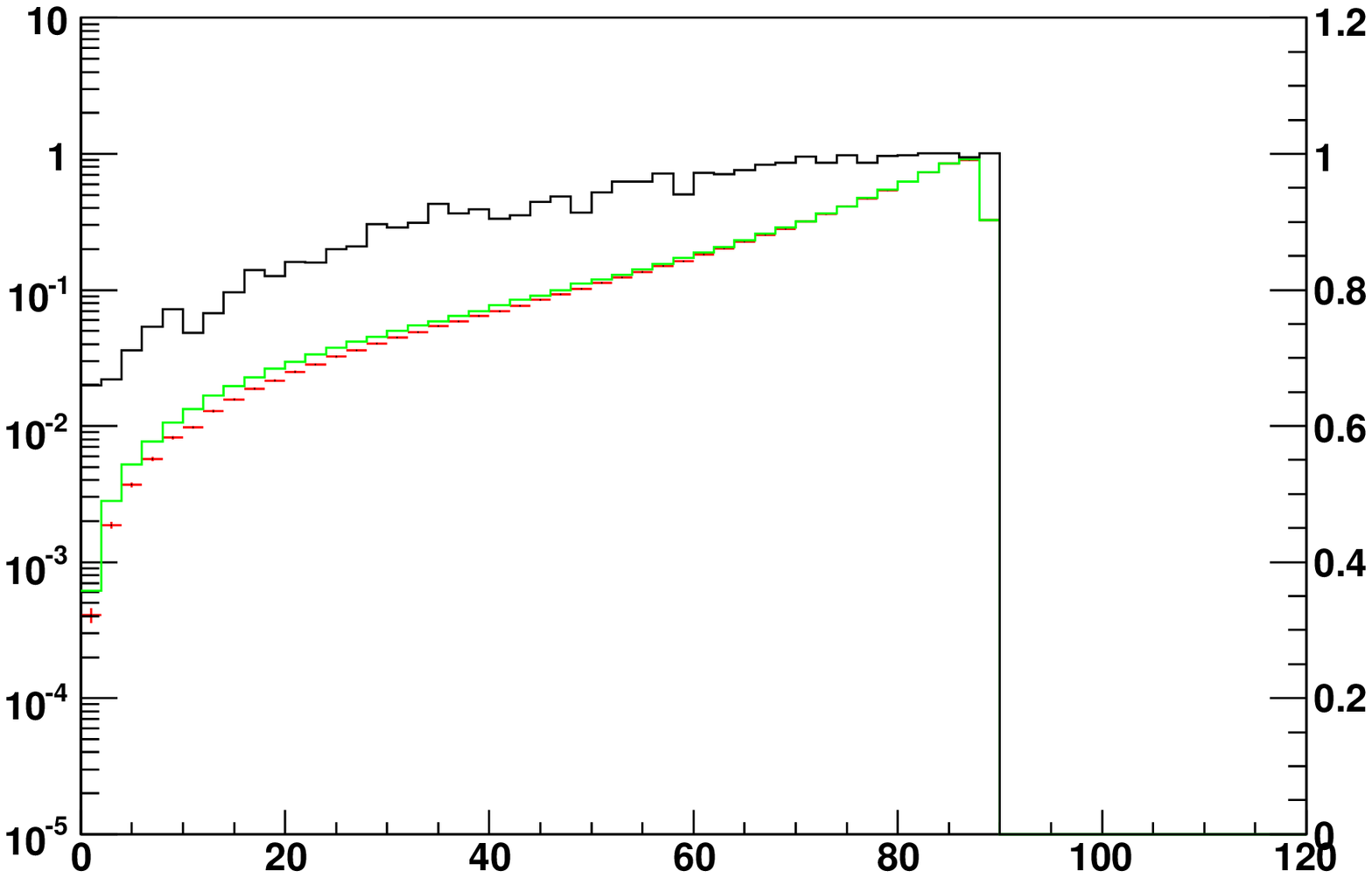}} }
{ \resizebox*{0.49\textwidth}{!}{\includegraphics{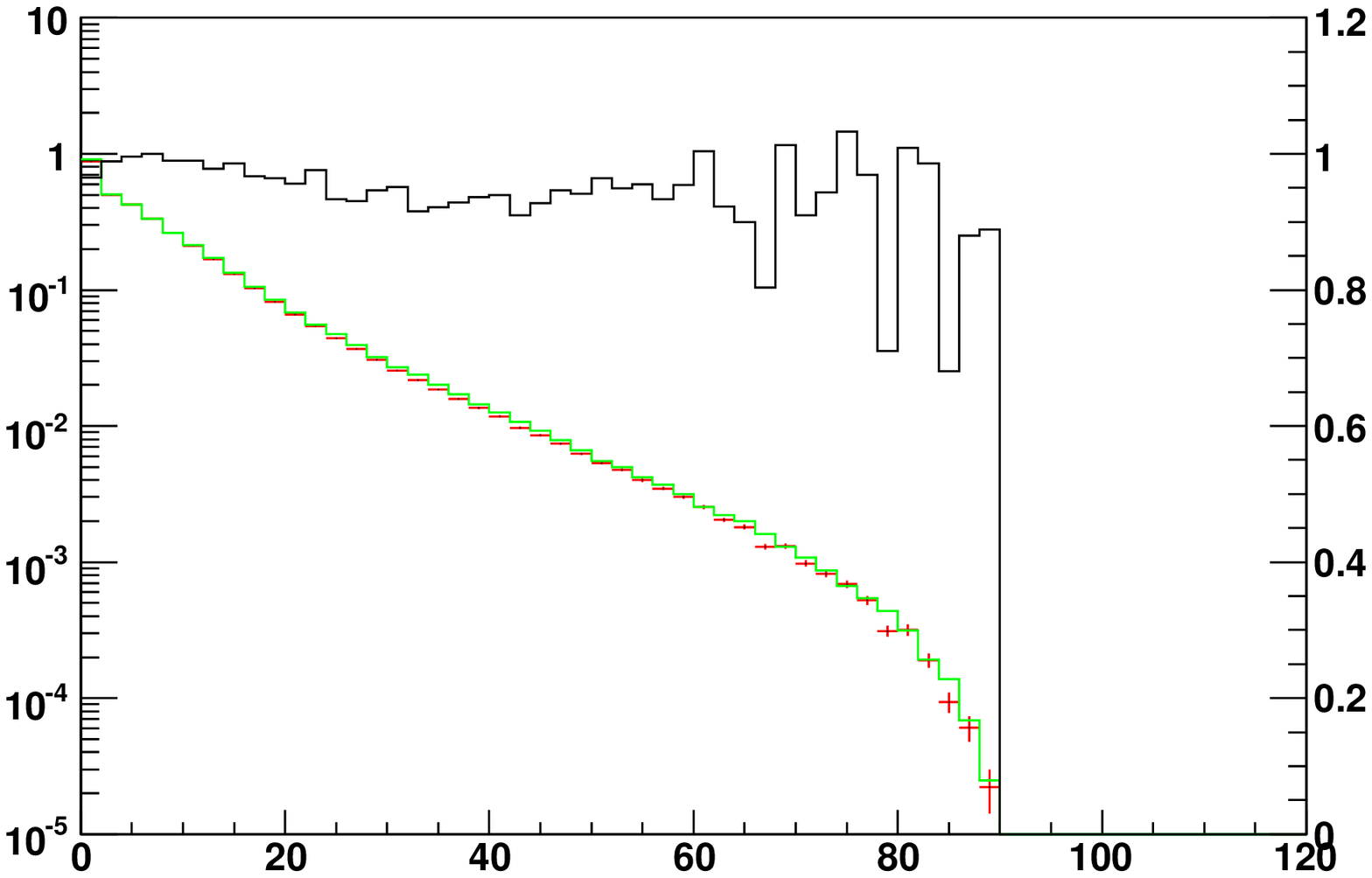}} }
\caption{
The comparison \cite{Golonka:2006tw} of the standard {\tt PHOTOS} (with multiple photon emission) and the {\tt KKMC} generator
(with second-order
matrix-element and exponentiation). In the left frame the invariant mass
of the $\mu^+\mu^-$ pair; SDP= 0.00918. In the right frame the invariant mass
of the  $\gamma \gamma$ pair; SDP=0.00268.
The fraction of events with two hard photons was
 1.2659 $\pm$  0.0011\%
for {\tt KORALZ} and
 1.2952 $\pm$  0.0011\%
for {\tt PHOTOS}. For the definition of shape difference parameter (SDP) 
see \cite{Golonka:2002rz}. \label{FigA}}}
\end{figure}

\begin{figure}
{\small
{ \resizebox*{0.49\textwidth}{!}{\includegraphics{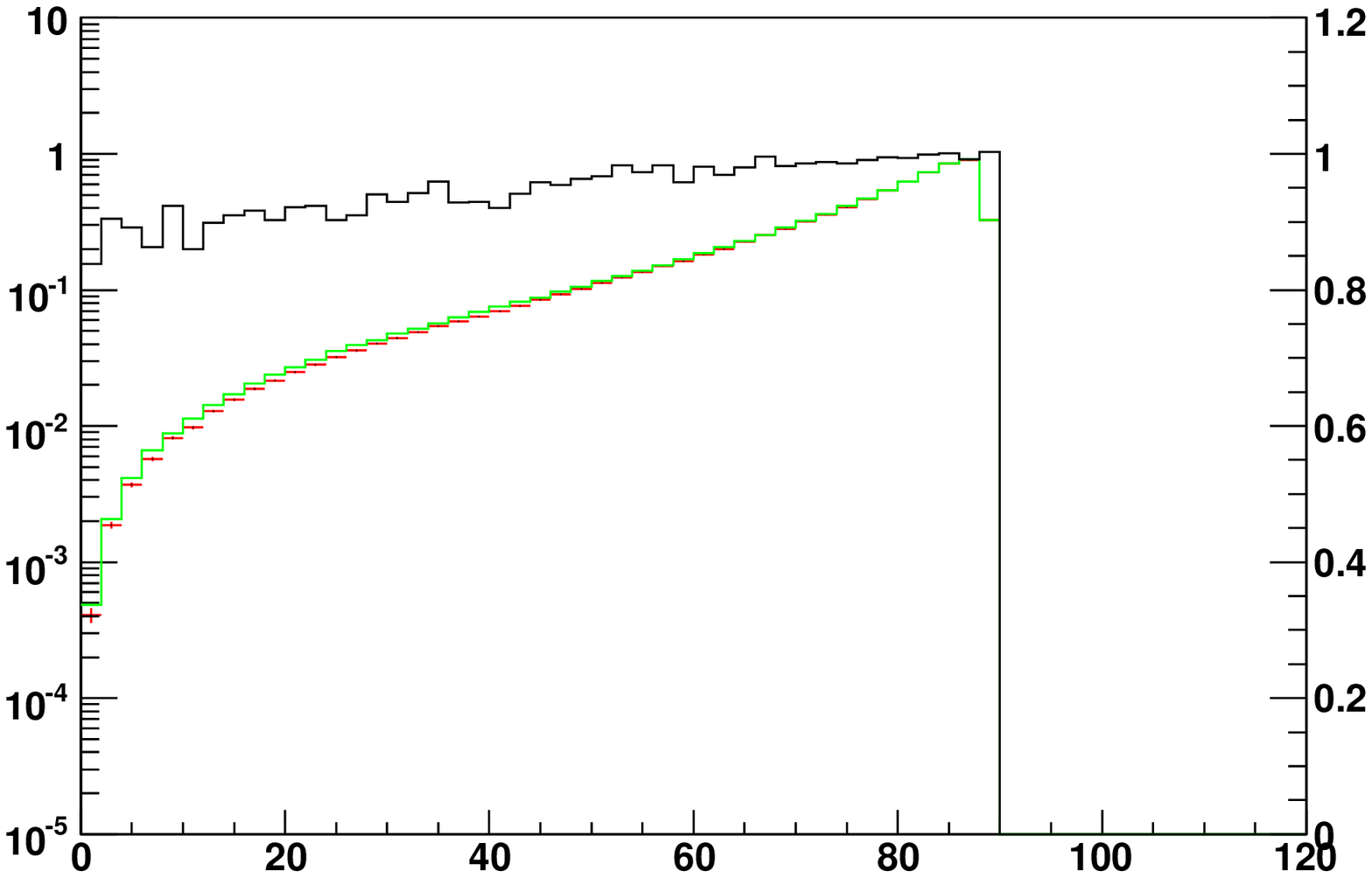}} }
{ \resizebox*{0.49\textwidth}{!}{\includegraphics{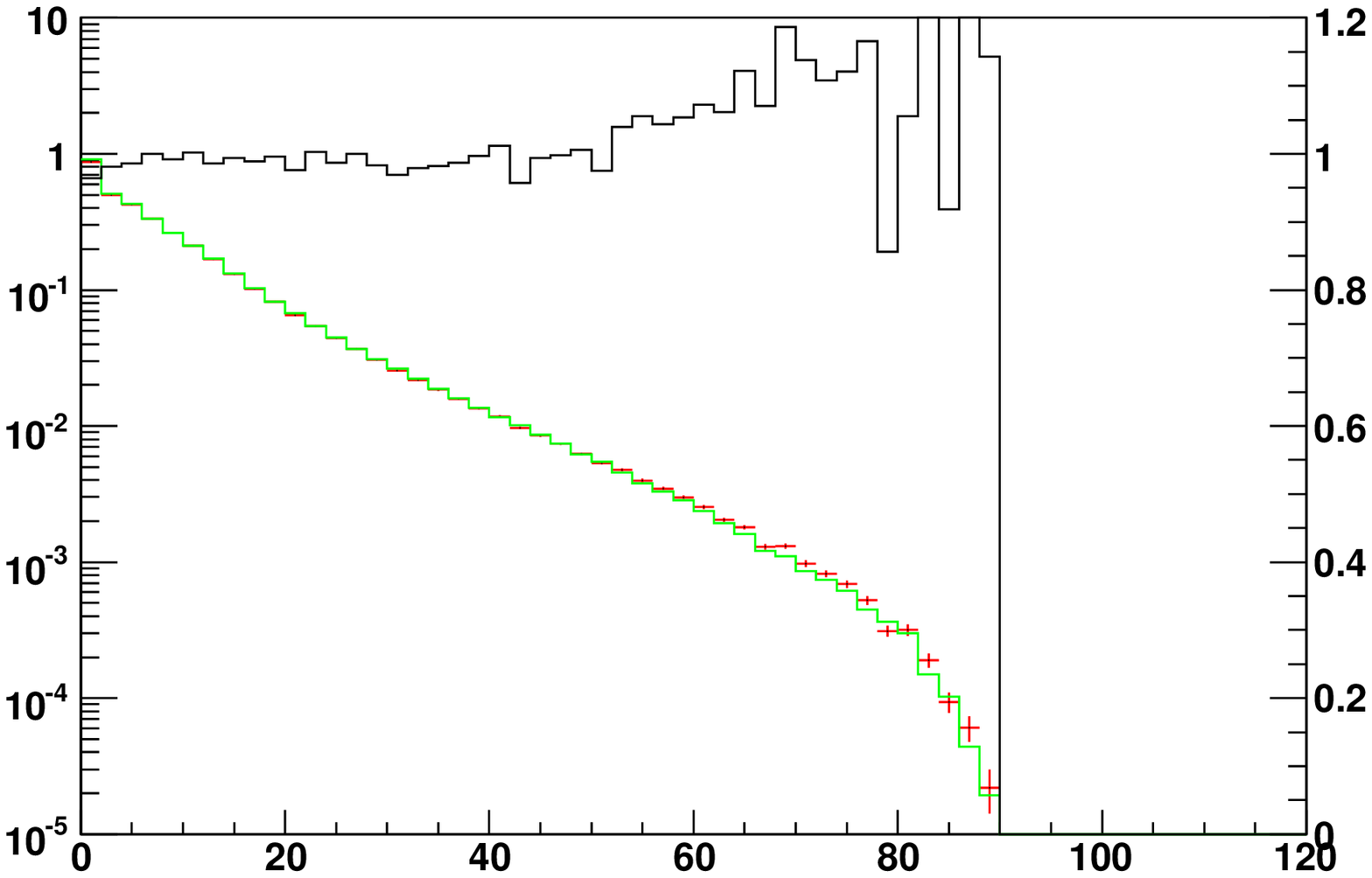}} }
\caption{{
The comparisons \cite{Golonka:2006tw} of the improved {\tt PHOTOS} (with multiple photon emission) and the {\tt KKMC}
generator ( with second order
matrix element and exponentiation).
In the left frame the invariant mass
of the $\mu^+\mu^-$ pair; SDP= 0.00142. In the right frame the invariant mass
of the  $\gamma \gamma$; SDP=0.00293.
The fraction of events with two hard photons was
 1.2659 $\pm$  0.0011\%
for {\tt KORALZ} and
  1.2868 $\pm$  0.0011\%
for {\tt PHOTOS}. For the definition of shape difference parameter (SDP) 
see \cite{Golonka:2002rz}. \label{FigB}}}}
\end{figure}

The NLO solution for {\tt PHOTOS}, as presented in ref.~\cite{Golonka:2006tw}, 
would therefore be feasible with some universal, {\it standard} event record,
nonetheless difficult due to practical issues of interfacing.
One should ask the question, what is the price related to the approximation
as implemented in public version of {\tt PHOTOS}.
 The results 
for this standard and NLO improved {\tt PHOTOS} are collected
in figures \ref{FigA} and \ref{FigB}. As one can see, improvement due to the use of exact 
first order matrix elements is unquestionable.
On the other hand, the  standard, easier to use, 
version seem to be sufficient in practically all phenomenological applications
as well.
 For the time being the problem of the optimal choice remains rather academic. 

\vspace{0.2cm}

\begin{figure}
 \begin{center}
   \includegraphics[ width=155mm,height=265mm, keepaspectratio]{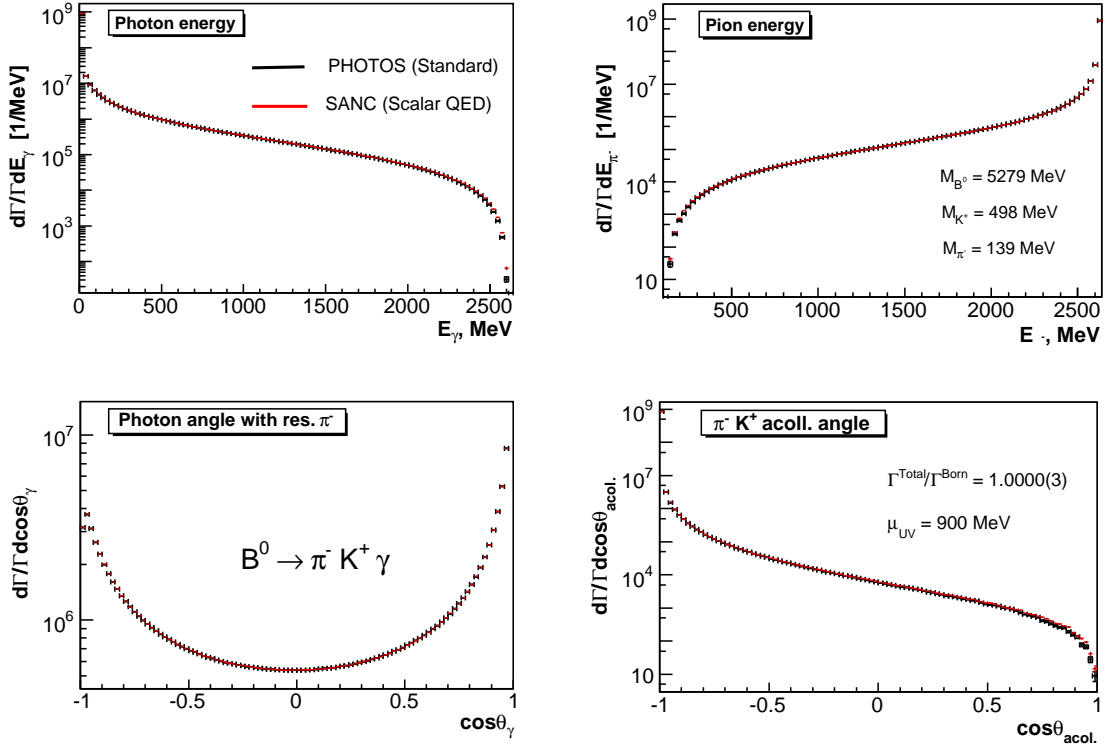}
 \end{center}
   \caption{\label{PmKp_distr_NotCorrected_p} Results \cite{Nanava:2006vv} from {\tt PHOTOS},
standard version, and {\tt SANC} for $B^0 \to \pi^- K^+(\gamma)$ decay are
superimposed on the consecutive plots. Standard
distributions, as defined in the text and logarithmic scales are used.
The distributions from the two programs overlap almost completely.
Samples of $10^9$ events were used.
The  ultraviolet scale, $\mu_{_{UV}}$, was chosen to leave total
decay width unchanged by QED.\label{FigC}
}
\end{figure}

\begin{figure}
 \begin{center}
  \includegraphics[ width=155mm,height=265mm, keepaspectratio]{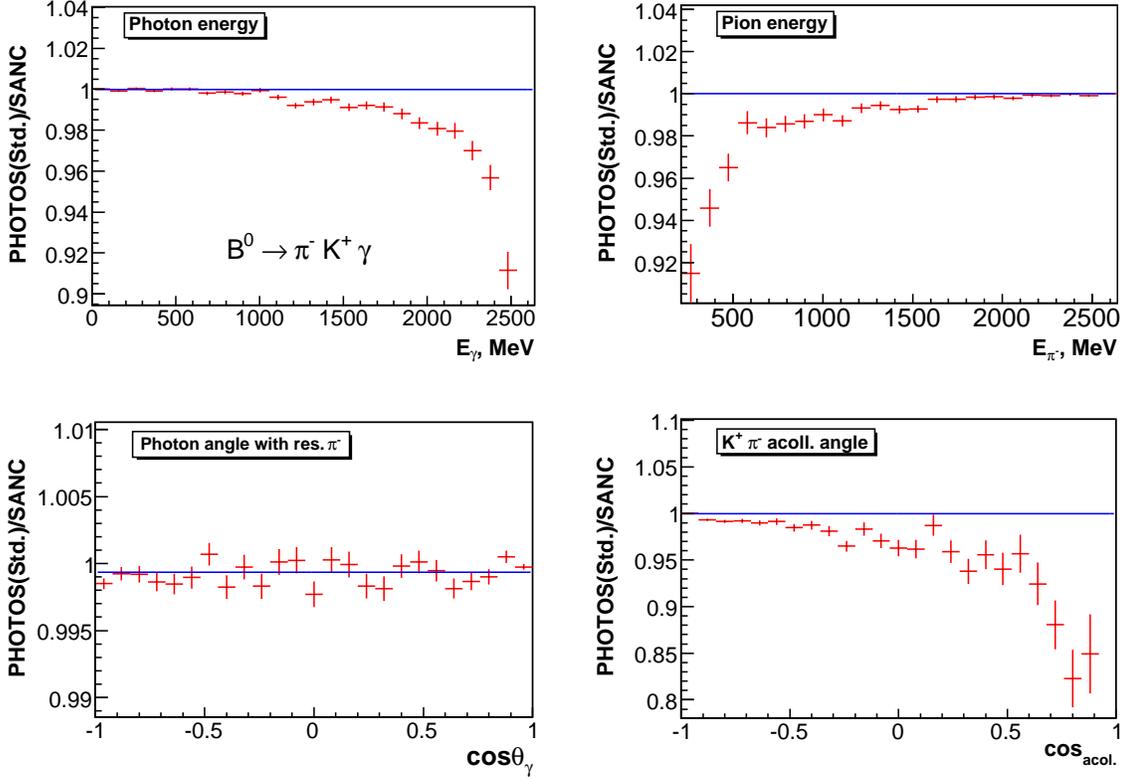}
 \end{center}
 \caption{\label{PmKp_ratio_NotCorrected_p} Results \cite{Nanava:2006vv} from 
{\tt PHOTOS},
standard version, and {\tt SANC} for ratios of the $B^0 \to \pi^- K^+(\gamma)$  distributions
are presented. Differences between {\tt PHOTOS} and {\tt SANC} are small, but are clearly visible now \label{FigD}
}
\end{figure}

\begin{figure}
\begin{center}
   \includegraphics[ width=155mm,height=265mm, keepaspectratio]{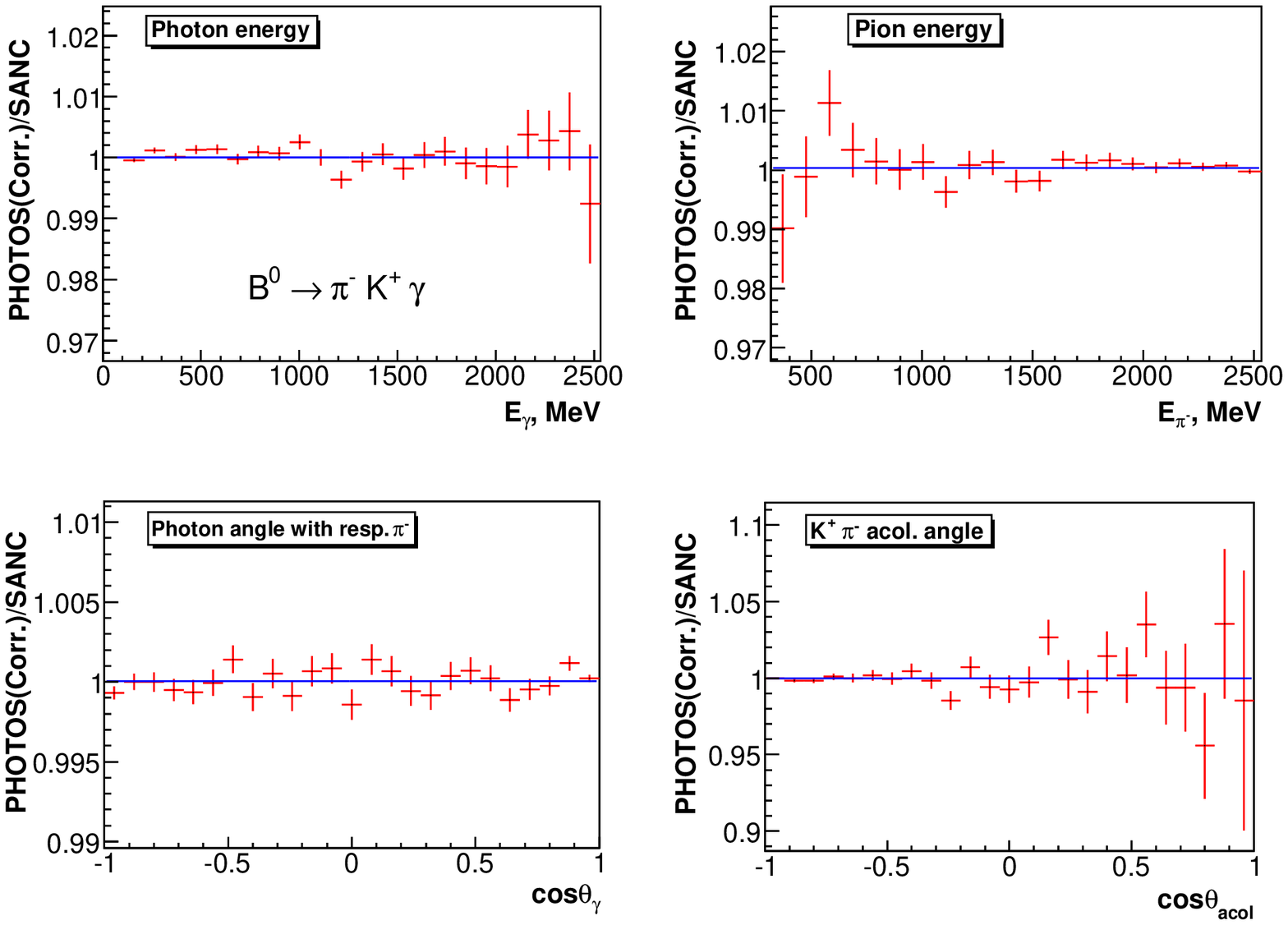}
\end{center}
\caption{\label{PmKp_ratio_corrected_p} Results \cite{Nanava:2006vv} from {\tt PHOTOS} with the exact matrix element,
and {\tt SANC}
 for ratios of the  $B^0 \to \pi^- K^+(\gamma)$ distributions. 
Differences between {\tt PHOTOS} and {\tt SANC} are below statistical error for samples of $10^9$ events.\label{FigE}
}
\end{figure}

In ref \cite{Nanava:2006vv}, we presented similar modifications in the {\tt PHOTOS} kernel
for the decay of $B$ mesons into a pair of scalars. As one can see from the comparison of plots
in figures \ref{FigC}, \ref{FigD} and \ref{FigE} the implementation of the exact (but scalar-QED only) kernel brings
a minuscule
improvement in the agreement between {\tt PHOTOS} and the reference exact simulation of
{\tt SANC} \cite{Andonov:2004hi}.
In this case both: {\tt SANC} and {\tt PHOTOS} are used to simulate single photon emission.
(There exists no reference simulation with which the multi-photon version of {\tt PHOTOS} could be
compared.)

For the NLO kernel in {\tt PHOTOS} the results are indistinguishable from those of
{\tt SANC}, even at statistical level of $10^9$ events. In this case, the 
price paid for improvement seems to be zero, as there is no need for extra information to be pumped from
the event record to the calculation of the {\tt PHOTOS} weight.
Actually, the exact kernel is even simpler than the standard one.

This high  precision as documented  in figs.~\ref{FigD} and ~\ref{FigE} 
is elusive: the dependencies on the production 
process may
appear if form-factors (originating from some unspecified here models) 
which have to be  fitted to the data.

From the technical side, one can interpret this excellent agreement as a strong test
of numerical performance of the program.
The necessary studies of the exact parametrization of the phase space used in {\tt PHOTOS},
which will also be important for future version of {\tt PHOTOS}, are described in detail in
the journal version of ref. \cite{Nanava:2006vv}.

\section{Mathematical aspects of the solution\label{SecSingProd}}

\begin{figure}
\begin{center}
                       \epsfig{file=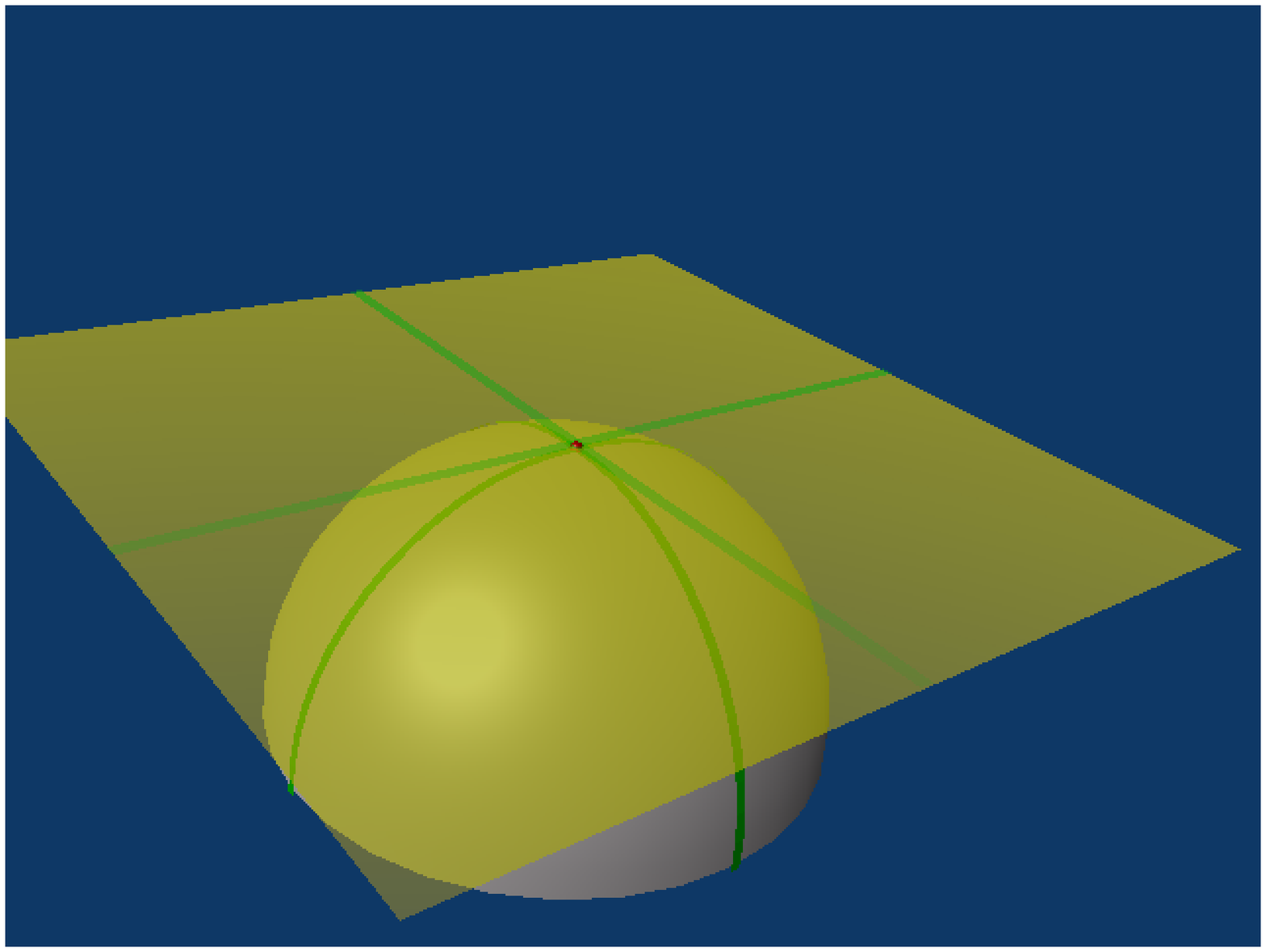,width=0.55\linewidth}
\caption{Symbolic reprentation of phase space with up to two extra particles. Curved surface 
represent actual phase space and the flat one tangent space. The thin bands represent 
configurations where only one extra particle is added.  Point in the center configuration 
of the Born level.  It is implicitly
assumed that particles of soft momenta do not provide much difference with respect
to configurations when they are absent. That is why symbolically such configuration seem to coincide.
\label{FigX}}
\end{center}
\end{figure}

One can ask if there is anything substantial in common in all these solutions
presented in \Section{SecNotation}, 
and whether systematization with the help of mathematical language is worth an effort.
Indeed, at the time of writing the first versions of the programs, which are now 
in a wide use, such considerations were  of low priority. In fact to a good
reason: they were expected to  slow progress and bring little.

At present, when multidude of different solutions is available and 
technical complexity of details dominates over main principles of construction
such effort may be well motivated and bring useful results.

\begin{figure}
\begin{center}
                       \epsfig{file=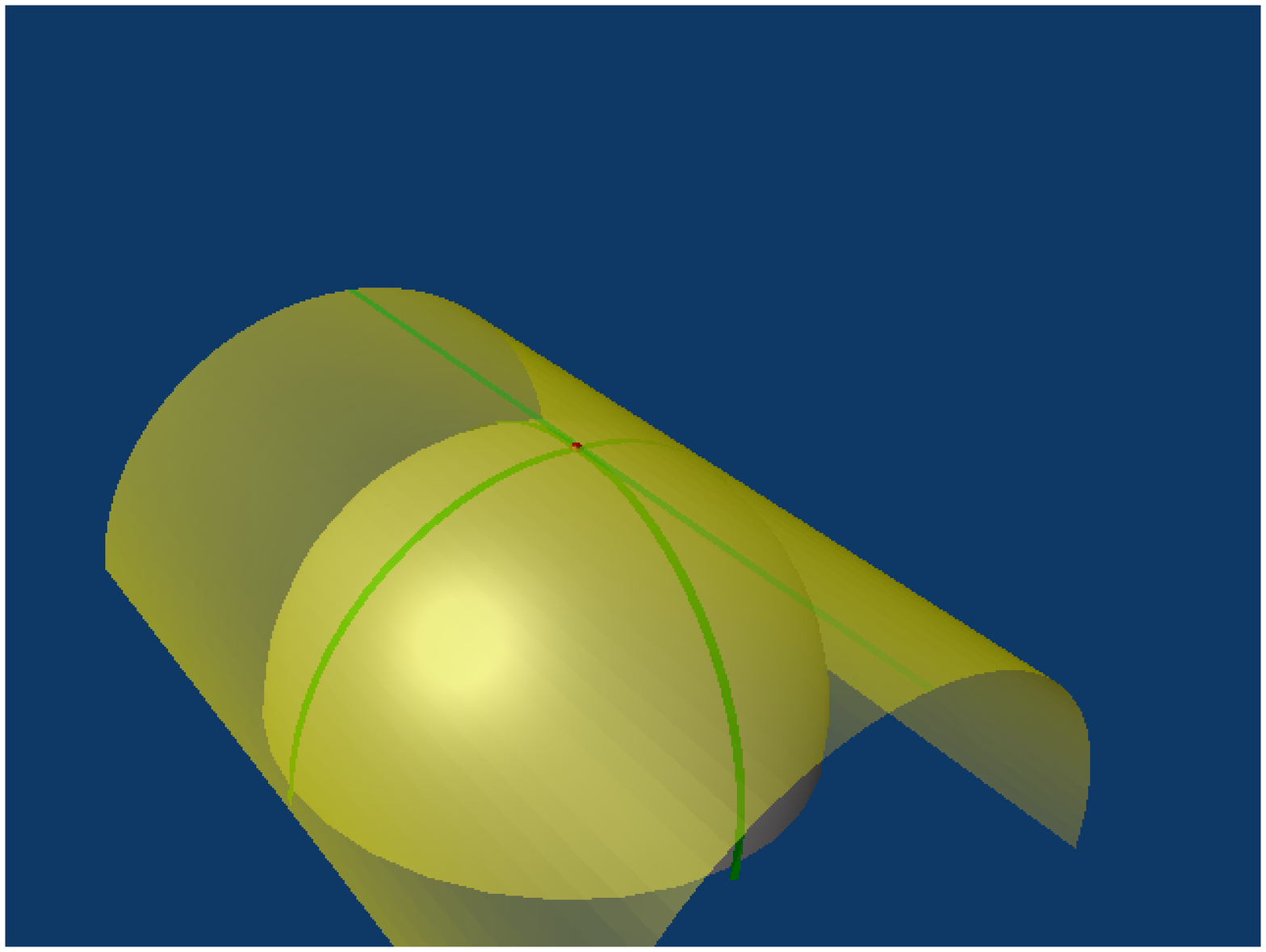,width=0.55\linewidth}
\caption{As in fig.\ref{FigX} this plot symbolically reprents phase space with up 
to two extra particles. Curved surface 
represent actual phase space and the cylindiric one the tangent space, where projection
of kinematical constraint of one of its dimensions was already executed.
\label{FigY}}
\end{center}
\end{figure}

Let us look at fig.~\ref{FigX} where 
 points, curved lines and surfaces on this heuristic plot 
represent consecutive manifolds of phase spaces for n, n+1, n+2 particles.
Note that the dimensionality of manifolds is in principle
counted by 
number of particles times dimension of Lorentz group representation, minus  overall
energy-momentum  and orientation constraints.
 Curvature appears as an  ultimate expansion parameter. The crude level 
distribution is also defined for phase spaces of n, n+1, n+2 particles
but as energy momentum constraint affects only first n particles, the further
ones
constitute flat Carthesian  sub-space. 
One step of the iterative projections 
as presented in Section~\ref{SecNotation} is symbolically presented in 
fig.~\ref{FigY}.

Case of   QED  and exponetiation of multiple photon radiation is 
rather simple, we do not need to worry about topological structure which is
the same for the final (physical)  phase space of multiphoton configuration and for 
the tangent space (constructed from eikonal phase space and matrix elements).
The projection 
from tangent space to the real one is trivial (at least from the point of view 
of topological properties).
In case of QCD we may expect complications, 
on the other hand hadonization models simplify the task anyway as they 
enforce separation of colour in the specific way. On the other hand it may   
be  unhelpful for the  discussion of the systematic errors.

There is another mathematical concept which is worth mentioning. 
Thanks to infrared sensitive regions of n+1 body phase space we obtain,
in a natural way, a triangulation line for this n+1 body phase space manifold.
In fact, structure of such induced triangulation  needs to 
be (topologically) the same for 
tangent and 
physical space, the  projections must match these triangulations.  
One can realize that the language of CW complexes (known in theory
of homotopy groups) 
may be useful to systematize the description and to separate it into easier to digest
parts.

Finally let us point to nice relation between {\tt PHOTOS}
algorithm for single (and fixed order) bremsstrahlung on one side and for 
the multibremsstrahlung cases. The relation
is a consequence of the properties of the tangent spaces.  It can be seen from
formal expansion of Poissonian distribution into sum of binomial ones.
 In the following formula we
identify coefficients of binomial and poissonian distributions:
 $p=\lambda$, $q=1-p$. Powers of $p$  denote  distinct
multiplicities.
\begin{eqnarray}
\exp(-\lambda) \sum_{n=0} \frac{1}{n!}\; p^n\; |_1 =
   & 1 \cdot (p+q)^1 \nonumber \\
\exp(-\lambda) \sum_{n=0} \frac{1}{n!} \; p^n\; |_2 =
   & \frac{1}{2} \cdot (p+q)^0 + \frac{1}{2} \cdot (p+q)^2\nonumber \\
\exp(-\lambda) \sum_{n=0} \frac{1}{n!} \; p^n\; |_3 =
   & \frac{2}{6} \cdot (p+q)^0 + \frac{3}{6} \cdot (p+q)^1 + \frac{1}{6} \cdot (p+q)^3\nonumber \\
\exp(-\lambda) \sum_{n=0} \frac{1}{n!} \; p^n\; |_4 =
  & \frac{9}{24} \cdot (p+q)^0 + \frac{8}{24} \cdot (p+q)^1 + \frac{6}{24} \cdot (p+q)^2 + \frac{1}{24} \cdot (p+q)^4
\label{expi}
\end{eqnarray}

These somewhat unexpected numerical constants, just ratios of natural numbers, 
provide trivial example  of expansion  of one set 
of special functions into another one. The consecutive lines of formula 
(\ref{expi}) correspond
to expansion at respectively $1^{st}$,$2^{nd}$,$3^{rd}$ and $4^{th}$ orders.
\section{Summary\label{SecSummary}}
%
In this talk 
we have presented some principles used in  Monte Carlo construction. It was a perfect
occasion to look into  history of projects, often common  with Prof. Jadach's.
For  that purpose iluminating mathematical aspects of the constructions seemed to be 
useful. They were one of the cornerstones in achieving quality and
robustness of the results. 
In the presented talk 
we have concentrated on phase space and its possible description with the
help of iterative Monte Carlo methods. Of course, main motivation of such a 
systematization  is to
search for prototypes of algorithms to be applied e.g. in QCD. 
Work on matrix elements was only marginally mentioned here. It is only starting, but some results 
could have been already presented now, see talk by  Andr\'e van Hameren. For more,
I am afraid, we need to wait for some time, even though some promising results 
are already available \cite{Jadach:2007qa,Placzek:2007xb}. The 
next anniversary Epiphany conference,
ten years from now, will hopefully bring some nice summary on that development.

\bibliographystyle{JHEP}
\addcontentsline{toc}{section}{\refname}\bibliography{ACAT07}

\end{document}